\documentclass[useAMS,usenatbib]{mn2e}
\usepackage{graphicx,epsfig}

\usepackage{enumerate}

\title[Do all QSOs have the same black hole mass?]{Do all QSOs have the same black hole mass?}
\author[T. Shanks et al]{T. Shanks$^{1}$\thanks{E-mail:
tom.shanks@durham.ac.uk (TS)}, S.M. Croom$^{2}$, S. Fine$^{1}$, N.P. Ross$^{3}$ \& U. Sawangwit$^{1}$\\
$^{1}$Dept. of Physics, Durham University, South Road, Durham, DH1 3LE, UK\\
$^{2}$School of Physics, University of Sydney, NSW 2006, Australia\\
$^{3}$Lawrence Berkeley National Laboratory, One Cyclotron Road, Berkeley, CA 94720, USA}
\begin{document}

\date{Accepted 2010 . Received 2010 ; in original form 2010 }

\pagerange{\pageref{firstpage}--\pageref{lastpage}} \pubyear{2010}

\maketitle

\label{firstpage}

\begin{abstract}
QSOs from SDSS, 2QZ and 2SLAQ covering an order of magnitude in
luminosity at fixed redshift exhibit similar amplitudes of clustering, with
the brightest sample showing a clustering length  only $11\pm9$\%
higher than the faintest sample. In addition, QSO clustering evolution
at $z>0.5$ is well fitted by a model that assumes a fixed host halo
mass. If halo and black-hole masses are related, then this may imply
that QSOs occur in a relatively narrow range of halo masses with a
correspondingly narrow range of BH mass. HST and Gemini high resolution
imaging of QSOs covering  a large range in luminosity also show a
relatively narrow range in QSO host galaxy luminosity. We argue that the
slow evolution of early-type galaxies out to $z\approx1-2$ may also
provide further support for a slow evolution of QSO host BH masses.  The
result would mean that if high-$z$ QSOs radiate at Eddington rates then
low-$z$ SyI must radiate at $\approx100\times$ less than Eddington. We
discuss the consequences in terms of four empirical models where (a)
QSOs radiate at a fixed fraction of $L_{Edd}$, (b) QSO luminosity
`flickers' over time, (c) QSOs have a single BH mass and (d) QSOs are
long-lived and evolve via Pure Luminosity Evolution (PLE). We conclude
that the $L_{Edd}$ model requires $M_{BH}$ and $M_{halo}$ to be
decoupled to circumvent the clustering results. While the single BH mass
and flickering models fit the $z>0.5$ clustering results, they appear to
be rejected by the $M_{BH}-L$ relation found from reverberation mapping
at $z\approx0$.  We find that the inclusion of $z<0.5$ QSO clustering
data improves the fit of a long-lived QSO model and suggest that the
predictions of the PLE model for QSO BH masses agree reasonably with
UV-bump and reverberation estimates.

\end{abstract}


\section{Introduction}

Over recent years there has been much new information on QSO clustering 
at high and low redshift. Here we combine results from various surveys to 
look specifically  for the luminosity dependence of QSO clustering.

The QSO clustering amplitude in the range $0.5<z<2.2$ evolves only
slowly with redshift. Since the mass clustering is falling as redshift
increases range, this means that the QSO bias appears to rise with
redshift. Converting bias into redshift via Press-Schechter theory
implies that, for example, 2QZ QSOs at most redshifts inhabit haloes of
$>2\times10^{12}M_\odot$\citep{croom05}. By comparing the space density
of QSOs with the space density of such haloes \citep{martini}, the
general conclusion is that QSOs are short lived, with lifetimes of
$\approx10^7$ yrs. This short lifetime has seemed a necessary
ingredient in a picture where the QSO halo mass does not appear to
increase with time under gravity \citep{croom05}.

QSO halo mass and BH mass are generally assumed to be correlated using
the empirical relations of \cite{ferrarese, gebhardt,ferrarese02,
wyithe} and we shall initially also be  making such an assumption. Some 
empirical support is found in the  observational analysis of
\cite{fine06} where BH masses estimated from BLR line widths are
compared to halo masses estimated from QSO clustering.  
The analyses of \cite{croom05}  suggest that it depends on
which of the above halo-BH mass relations are assumed as to whether
there is evolution in the Eddington ratio with redshift or not.
Certainly the errors are large enough to encompass various
possibilities. 

\citet{croom05} found little evidence for luminosity dependent QSO
clustering at fixed $z$ in the 2QZ survey, however the luminosity range
in any individual magnitude limited survey is necessarily limited. So
\citet{pn06} and \citet{shen09} claimed some evidence for luminosity
dependent QSO clustering respectively within the individual 2QZ and SDSS
QSO surveys but both results were only marginally significant. Then
\citet{jose08} found little evidence of a luminosity dependence of
clustering after adding the fainter 2SLAQ survey to the 2QZ survey. They
suggested that the results might be in good agreement with the models of
\citet{lidz06} where the results of \citet{croom05} were interpreted as
suggesting that if QSOs have a short lifetime and evolve quickly from
bright to faint states (`flickering'), then it might be expected that
the luminosity dependence of QSO clustering might be small as found by
\citet{croom05}.

Here we shall first find further evidence for the
independence of QSO clustering and luminosity supporting these previous
conclusions. We shall then consider the implications for QSO evolution
by considering  four simple empirical models where (a) QSOs radiate at a
fixed fraction of $L_{Edd}$ (eg \citealt{rees84}), (b) QSO luminosity
`flickers' over time \citep{lidz06}, (c) QSOs have a single BH mass
\citep{croom05,jose08} and (d) QSOs are long-lived and follow Pure
Luminosity Evolution (PLE, \citealt{bsp87, boyle88}). We shall attempt to
test each of these against the clustering and other QSO data. In the
cases where models fail the QSO clustering test, we shall also consider 
whether the  assumptions that underpin the clustering test may be at fault 
rather than the model.

The data that we will use comes from  the SDSS DR5, 2QZ and 2SLAQ QSO
redshift surveys. These surveys provide a range of magnitude limits
which cover a factor of 10 in QSO luminosity. This means we can for the
first time efficiently check the dependence of small-scale clustering on
QSO luminosity. We shall use these results as a context for a discussion
on the phenomenology of QSO formation and evolution.

\section{QSO Clustering data}
\subsection{SDSS, 2QZ and 2SLAQ  surveys}
Previously \citet{croom05} used the 2QZ survey to estimate the QSO
correlation function and its dependence on redshift and luminosity. This
was a survey of $\approx22655$ QSOs in two $\approx375$deg$^2$ strips in
the NGC and SGC. The magnitude limit was $18.25<b_J<20.85$ and the
resulting QSO sky density was $31$deg$^{-2}$. At the average redshift of
$z=1.4$, the average absolute magnitude is $M_{b_J}\approx-25$mag.
\cite{croom05} measured $s_0=5.4^{+0.42}_{-0.48}h^{-1}$Mpc and
$\gamma=1.2\pm0.1$ in the range $1<s<25$h$^{-1}$Mpc for the amplitude
and slope of the redshift-space  correlation function, $\xi(s)$.

\citet{jose08} then used the 2SLAQ survey of 9418 QSOs based on SDSS
imaging to test the luminosity dependence of the QSO clustering. The
magnitude limit was $20.85<g_{AB}(\approx b_J)<21.85$ with a sky density
of the resulting QSO sky density was $\approx49$deg$^{-2}$, including
the 2dF QSOs where the two surveys overlapped, in a total area of
192deg$^2$. The average absolute magnitude at $z=1.4$ is
$M_{b_J}\approx-23.5$mag. \citet{jose08} found a $\xi(s)$ amplitude and
slope similar to that for 2QZ and again found that this QSO clustering
amplitude depended very little on redshift or luminosity.

Most recently, \citet{ross09} have analysed the clustering of 30239 QSOs
in the 5740deg$^2$ SDSS DR5 survey to a limit of $i_{AB}=19.1$,
producing an average absolute magnitude of $M_i\approx-26.3$mag or
$M_{b_J}\approx-26.1$ at $z\approx1.4$. This provides a sky density of
5.3deg$^{-2}$ in the uniform sample where \citet{ross09} measured
$s_0=5.95\pm0.45h^{-1}$Mpc and $\gamma=1.16_{-0.16}^{+0.11}$ in the 
$1<s<25$h$^{-1}$Mpc range. 

\subsection{Clustering comparison}
In Fig. 1   we now compare the clustering results from the three surveys
directly with each other in terms of the redshift-space correlation
function, $\xi(s)$. The cosmology assumed  in all cases is
$\Omega_\Lambda=0.7$, $\Omega_m=0.3$. We see that all three results are
quite consistent, particularly in the range 5-20h$^{-1}$Mpc where the
signal is excellent. The power-law amplitudes and slopes the authors
have fitted are also consistent. We checked this by fitting the $\xi(s)$
results consistently in the $1<s<30$h$^{-1}$Mpc range. We have fitted
for the real space correlation function scale-length, $r_0$, assuming
power-law slope, $\gamma=1.8$, infall parameter, $\beta=0.4$, and
line-of-sight pairwise velocity dispersion $<w^2>^{1/2}=750$kms$^{-1}$
The latter, somewhat degenerate, parameters are in the range usually
found in redshift-space distortion analyses of QSO surveys
\citep{hoyle02,jose08}. We find good  consistency in these results with
SDSS giving $r_0=6.30\pm0.3$h$^{-1}$Mpc, 2QZ giving
$r_0=5.75\pm0.25$h$^{-1}$Mpc and 2SLAQ giving
$r_0=5.70\pm0.35$h$^{-1}$Mpc. The best overall fit is
$r_0=5.90\pm0.14$h$^{-1}$Mpc. The fits are simple $\chi^2$ fits, with no
account taken of covariance between the correlation function points,
since previous results have shown that the sparse QSO sampling leaves
them approximately independent, at least at the bin sizes we have used
\citep{croom05,jose08}. The brightest QSO sample scale length (SDSS) is
therefore only $11\pm9$\% larger than the faintest sample. The overall
$\chi^2$ goodness of fit of all 3 surveys to the
$r_0=5.90\pm0.18$h$^{-1}$Mpc model is 46.2 on 36 degrees-of-freedom,
acceptable at the $1.75\sigma$ level. Thus in surveys ranging from
magnitude 19.1 to 21.85 or a factor of $\approx10$ in luminosity
($-26.1<M_{b_J}<-23.5$), the QSO clustering amplitudes are remarkably
similar. Since the three surveys have very similar redshift
distributions, this means that the amplitude of clustering at fixed
redshift ($z\approx1.4$) is approximately independent of QSO luminosity.
We further note that other authors have measured the cross-clustering of
low redshift QSOs with LRGs and have found a similar lack of dependence
on luminosity. For example, \citet{gm} found that $z\approx0.7$ LRGs
cross-correlated with SDSS, 2QZ and 2SLAQ QSOs with similar amplitudes.
Thus both auto- and cross-correlation analyses support the remarkable
idea that intrinsically bright and faint QSOs show similar clustering
environments.

\begin{figure}
\includegraphics[scale=0.5]{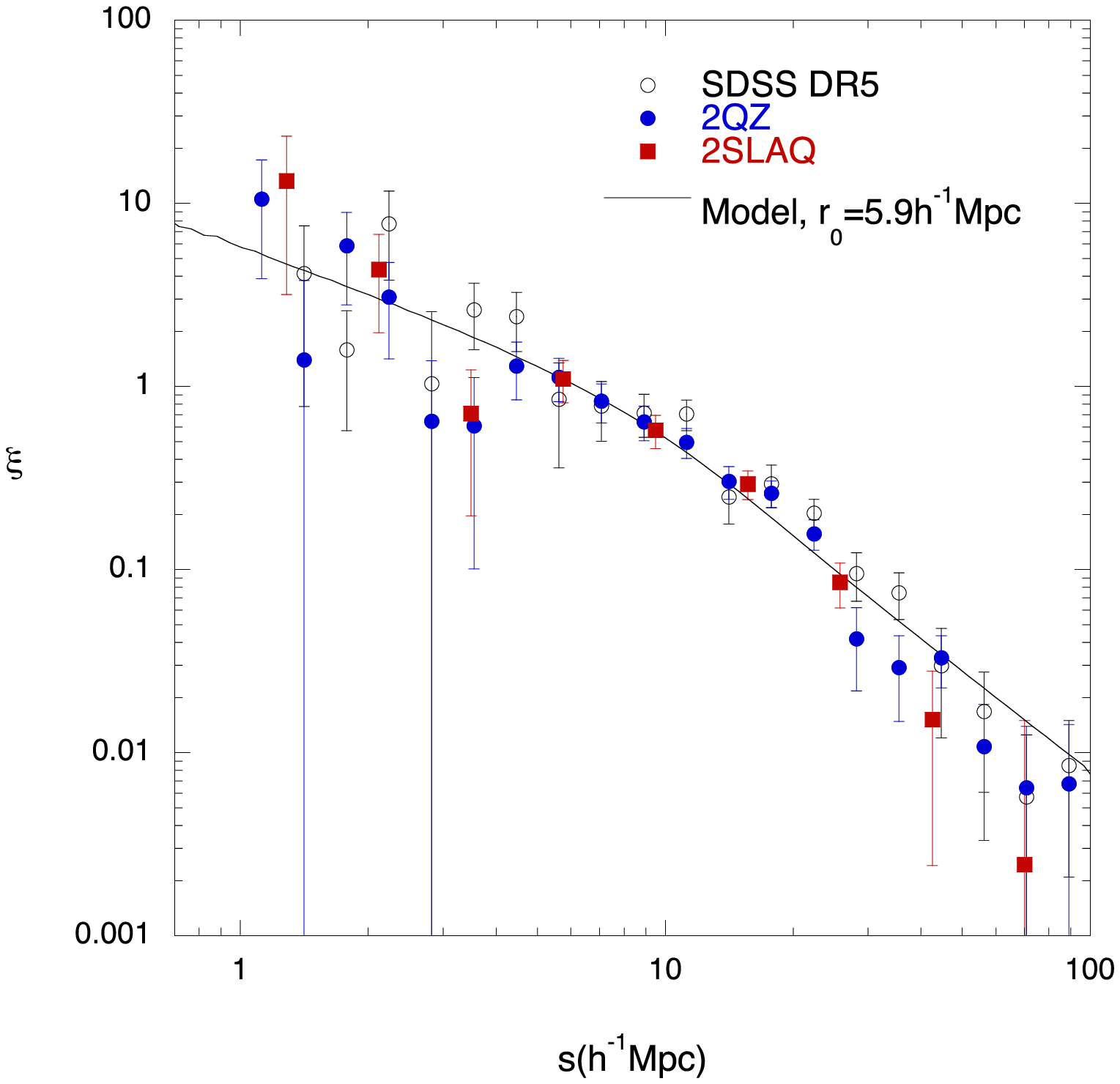} 
\caption{The QSO correlation function $\xi(s)$ compared for the SDSS,
2QZ and 2SLAQ surveys.These surveys span a  factor of $\approx10$ in
luminosity at fixed redshift ($z\approx1.4$) but the clustering
amplitudes are very similar. The model assumes $\xi(r)=(r/r_0)^{-1.8}$
with line-of-sight velocity dispersion $<w^2>^{1/2}=750$kms$^{-1}$ and
infall parameter $\beta(z=1.4)=0.4$. The assumed cosmology is
$\Omega_m=0.3$ and $\Omega_\Lambda=0.7$.}
\label{fig:xis}
\end{figure}

\section{Halo and BH masses via clustering evolution}

\begin{figure}
\includegraphics[scale=0.5]{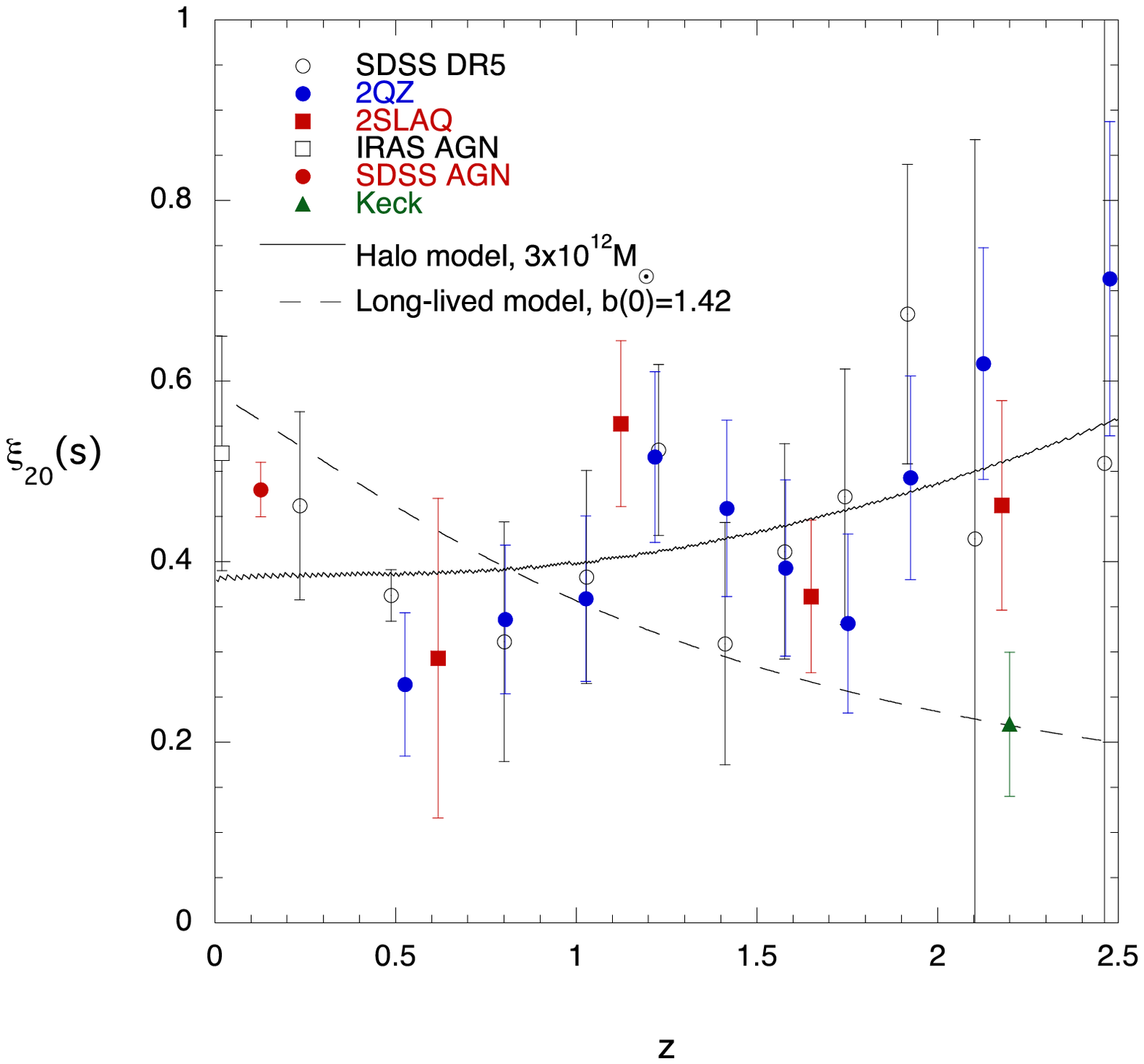}
\includegraphics[scale=0.5]{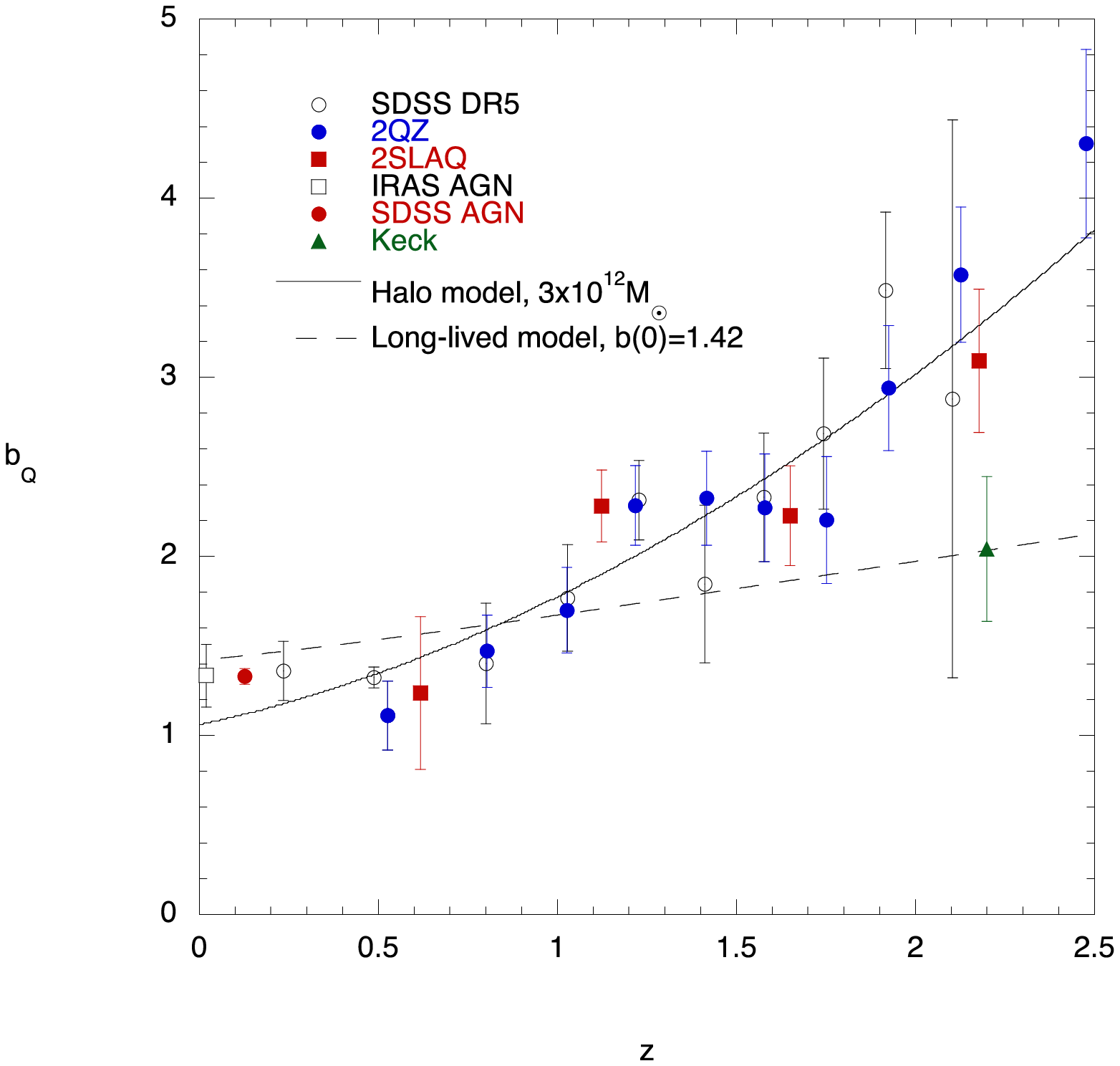}
\caption{(a) $\xi_{20}$ versus redshift for the SDSS, 2QZ and 2SLAQ QSO
surveys, for an assumed $\Omega_\Lambda=0.7$, $\Omega_m=0.3$ cosmology.
Also shown are the results for the SDSS AGN of \citet{wake04}, the IRAS
AGN of \citet{georg} and the Keck QSO-LBG cross-correlation result of
\citet{steidel} with $\xi_{20}$ as reported by \citet{jose08}. The solid
line represents the model for fixed  halo mass of
$3\times10^{12}M_\odot$. The dashed line represents the evolution of the
long-lived model of \citet{fry96} which best fits the data with
$b_Q(z=0)=1.42\pm0.03$. (b) As above for the linear QSO bias, $b_Q$, now 
calculated from the real-space QSO correlation functions.}
\label{fig:xi20_lam}
\end{figure}

We next consider the halo model fit  to the bias-redshift relations
shown  taken from  \citet{croom05} for 2QZ, \citet{jose08} for the 2SLAQ
survey and  \citet{ross09} for the SDSS sample. These results have shown
 that the bias at each redshift is compatible assuming linear
fluctuations and  Gaussian density distributions with the QSOs occupying
a single halo mass of $\approx3\times10^{12}M_\odot$ at all redshifts.
If we also assume that there is a host halo mass - BH mass relation (e.g
\citealt{ferrarese02}), this means that QSOs may contain the same
BH-mass at all redshifts. A model with fixed halo mass of
$M_{halo}=3.0\pm0.38\times10^{12}M_{\odot}$ as found by \citet{croom05}
for 2QZ is compared to the evolution of $\xi_{20}$ in Fig.
\ref{fig:xi20_lam}a and to the overall QSO bias-redshift relation in
\ref{fig:xi20_lam}b. The normalisation for the mass
model is also chosen to be the same as that of \citet{croom05}. A fit
based on the SDSS, 2QZ and 2SLAQ QSO datasets gives a  fitted halo mass
of $M_{halo}=3.27\pm0.41\times10^{12}M_{\odot}$ with
$rms=2.0\times10^{12}M_{\odot}$, consistent with the result for 2QZ. It
can be seen that the single halo mass model is a very good fit to these
data, although we note that this model does not represent the evolution
of an individual QSO. Following \citet{croom05}, their equation (24)
which assumes the unevolving relation, $M_{BH}\approx M_{halo}^{1.82}$
from \citet{ferrarese02} then gives  $M_{BH}\approx5\times10^8M_\odot$,
again approximately independent of redshift. We note that the model
dependence of  this mass is large with the different  $M_{BH}-M_{DMH}$
models of \citet{ferrarese02} and the evolving model of \citet{wyithe}
predicting masses between $10^8M_\odot$ and $10^{10}M_\odot$ at fixed
$z$ \citep{croom05}. Nevertheless the relative $M_{BH}$  change  with
$z$ is much less model dependent and this is more important for our
purposes here.

These conclusions only apply if we  base our results on the SDSS, 2QZ
and 2SLAQ samples. Including the IRAS AGN, SDSS AGN and Keck points only
increases the result to $M_{halo}=3.74\pm0.57\times10^{12}M_{\odot}$ but
the single halo mass model is now marginally rejected by the data at the
2-3$\sigma$level ($\chi^2=42.8$ on 26 dof, $P<0.025$), with the highest
residuals coming from the SDSS AGN and Keck points (see Figs.
\ref{fig:xi20_lam}) and there is evidence for an increase in halo and
hence black-hole masses as we move to lower redshift. For example, the
halo mass corresponding to the SDSS AGN sample at $z\approx0.13$ is
$M_{halo}=9.2\pm1.7\times10^{12}M_\odot$. It might be argued that low
redshift Seyfert 2 galaxies may be a different class from the Type I
QSOs that dominate at higher redshift but given their lower
luminosities, it might be expected that their clustering amplitude
represents a lower limit to that of Seyfert I's at low redshift.  In a
unified picture these AGN might be expected also to be representative of
Seyfert I's. One caveat is that the SDSS AGN sample may also contain a
population of LINERS which may not be comparable to Seyfert I in their
clustering properties. Later we shall also argue that their
$\approx20\times$ higher space density means that only a fraction are
expected to be obscured Seyfert I's. We also note that \cite{hickox} has
suggested that obscured QSOs, with $r_0=6.0\pm0.6$h$^{-1}$Mpc, may show
a higher  clustering amplitude than unobscured QSOs, with
$r_0=5.3\pm0.6$h$^{-1}$Mpc, in their QSO samples. We postpone further
discussion of these two lower redshift clustering points until Sections
7 and 8.

\section{QSO host galaxies via direct imaging} 

We now look at the empirical evidence for the  mass and luminosity of
QSO hosts from high resolution direct imaging. \citet{schade} used HST
to image X-ray QSOs out to $z\approx0.5$, where  imaging decomposition
is most reliable. They combined these data with those from similar
observations of higher luminosity QSOs from \citet{bahcall97} and
\citet{boyce98}. Fig. 13c from these authors shows a flat distribution
of host galaxy luminosity with QSO luminosity. This is particularly the
case considering that  low luminosity hosts of high luminosity QSOs are
particularly difficult to detect. These authors also note that when a
further decomposition into a bulge and disk is performed, there may then
be more dependence of host bulge luminosity on QSO luminosity. The
errors on the bulge luminosity become increasingly large and may provide
an explanation for the wider scatter that is seen in this relation.
There are also observations (e.g. \citealt{merloni}) and model
predictions (e.g. \citealt{lamastra}) suggesting  that the bulge-BH mass
relation may not apply at high redshift (see also \citealt{mclure06}).
The potentially different results for bulge and galaxy luminosities must
also be borne in mind because luminosity or more accurately disc stellar
mass may better correlated with DM halo mass than bulge mass (e.g
\citealt{korm11a,korm11b}).

This work was extended to higher redshift by \citet{croom04mn} using
Gemini ALTAIR$+$NIRI observations of 10 luminous QSOs out to
$z\approx2$. Only one host galaxy was detected and the upper limits on
the others were compatible with simple passive evolution models of the
host from the present day. Models where the hosts evolved at the same
rate as the QSO luminosities were rejected at high significance. Again
if we assume that the bulge-BH mass relation applies at high redshift
then these results continue to imply a constant BH mass for high-z QSOs.
Similar results have also been reported by other authors. For example,
\citet{floyd}, in HST observations, also found only an $\approx 2$ mag
range of galaxy host luminosity for  an $\approx 4-5$mag range of QSO
luminosity, with little evidence for a correlation between the two,
except possibly in their upper envelope (see their Fig. 7, upper panel). Thus the
results from QSO clustering analyses and from QSO direct imaging  appear
to support the suggestion that QSO host galaxies may have a luminosity or,
by extrapolation, stellar mass that is independent of QSO luminosity.

\section{Immediate  implications for QSO models}

Based on the SDSS, 2QZ and 2SLAQ samples, we have found  that either the
clustering method is not appropriate or that all QSOs dwell in similar
sized haloes. This could further  imply that the broad-lined QSO
population contain BHs of similar mass, approximately independent of QSO
luminosity. The clustering results are powerful with the capability to
rule out simple models. For example, the results appear incompatible
with the simple model where QSOs all radiate at a fixed fraction of
$L_{Edd}$, unless there is no relation between $M_{BH}$ and $M_{halo}$.
Many other models that make this underlying assumption will also have
this problem with these data. For example, Fig. 10 of \citet{bonoli09}
predicts that at $z=1.5$, $M_B<-24$ QSOs give $r_0=7$h$^{-1}$Mpc   and
$M_B<-22.5$ QSOs give $r_0=5.25$h$^{-1}$Mpc, ie a difference of $\Delta
r_0=1.75$h$^{-1}$Mpc over this  $\approx1.5$ mag luminosity range. In
comparison, over a wider 2.75 mag range, the data  in Fig. \ref{fig:xis}
only gives  $\Delta r_0=0.6\pm0.46$h$^{-1}$Mpc. Thus  the predicted
$\Delta r_0=1.75$h$^{-1}$Mpc is marginally  rejected at $2.5\sigma$. (We
ignore the error on the model at fixed $z$ because the predicted
luminosity dependence clearly persists at all $z$.) Similar conclusions
apply to the simplest version of the semi-analytic model of
\citet{fanidakis} for the evolution of black hole growth. Other
empirical models such as the QSO `flickering' model of \citet{lidz06}
and the single BH mass QSO model of \citet{croom05,jose08} do better,
since they were designed to explain the early indications of the
luminosity independence of QSO clustering.

Thus, QSO host halo and BH  masses at fixed $L/L^*$ may be independent of 
redshift despite the fact that the QSO luminosity brightens by a factor
of $\approx30$  in the range $0<z<2.2$. Moreover, since  QSOs at fixed
redshift have the same clustering strength independent of luminosity or
$L/L^*$, this could imply that {\it all} QSOs have a narrow distribution
of BH mass, independent of both redshift and luminosity, with  QSOs
becoming increasingly sub-Eddington at lower luminosities.

This appears unlikely. However there are some pieces of evidence that
this might be the case. First, the luminosity function of the early-type
galaxies thought to host QSOs appear to evolve very little over the
redshift range out to $z\approx1-2$ \citep{nm06, wake06, brown07}. If
the LF and stellar mass function do not evolve much then it is a short
step to supposing that the dynamics of the early-type do not evolve much
either (although see \citealt{daddi05}). Then it could be that the BH mass
might also be unevolving. The biggest BH lie in the biggest bulges and
this relation is then predicted not to evolve with time. The QSOs then
have to inhabit a narrow range of bulge/BH mass but with a large range
of luminosity at fixed $z$. 

An escape from this conclusion could be that the high-$z$ QSOs have
moved along the galaxy luminosity/mass function to higher galaxy
luminosities/masses  and hence have higher BH masses more consistent
with being at Eddington. However, the fact that it has been so difficult
to identify QSO hosts at $z\approx2$ in high resolution imaging led
\cite{croom04mn} to suggest that the data are consistent with the hosts
being similar luminosity to those at lower redshift.

Another possible argument for a slowly changing BH mass with time is the fact
that the velocity widths, $<~v^2>^{1/2}$, of the broad lines are similar
for high luminosity QSOs at high $z$ as they are for low-luminosity
Seyfert I's at low $z$. Now in the standard view where QSOs radiate at
approximately a fixed fraction of the Eddington luminosity, this is just
a coincidence where the increase in BH mass by a factor of $\approx30$ at
high $z$  is approximately offset by a comparable increase in the BLR
radius. Since the BLR radius-luminosity relation takes the form
$r \approx L^{0.5-0.7}$ \citep{bmp04,kaspi05,bentz09}, if  this relation
is unevolving with redshift then $<v^2> \propto L/r \propto L^{0.5-0.3}$
and so $<v^2>^{1/2} \propto L^{0.25-0.15}$, perhaps slow enough to
explain the similarity in line widths. In this fixed fraction of
$L_{Edd}$ case, there is therefore some empirical evidence to help
explain this coincidence. Nevertheless, the explanation is still
vulnerable if the $r-L$ relation actually proves to evolve with redshift.

Alternatively, to maintain the single BH mass model, we could  pursue
the idea motivated by the early-type galaxies that nothing evolves
except the QSO luminosity. On this view, the BLR radius  might then
remain constant with luminosity and hence redshift ie $r\approx L^0$. A
fixed broad-line velocity width, $<v^2>^{1/2}$, and radius, $r$, gives a
fixed BH mass with luminosity/redshift. The question then is whether it
is feasible, for example, that the factor of $\approx30$ increase in
$L^*$ over $0<z<2.2$ could leave the BLR radius unchanged if a balance
has to be maintained between photoionisation and gravity and  this seems
physically unlikely. It also seems at odds with the above evidence for
the $r-L$ (and $M_{BH}-L$) correlation found from reverberation mapping.
Thus if these empirical results are correct then the constancy of QSO
velocity width with $L$ and $z$ must remain simply a coincidence, rather
than providing further evidence for the single BH mass model.

Indeed, any single BH mass/flickering  model motivated by QSO clustering 
clearly has a severe problem with the QSO $M_{BH}-L$ relation
found from reverberation mapping at low redshift. Equn (9) of
\citet{bmp04} rejects no correlation ie a slope of zero  in the
$M_{BH}-L$ relation at the $8.8\sigma$ level! So here we have
potentially the most serious and direct  conflict with any simple single
black hole mass interpretation of the QSO clustering results. As
\citet{bmp04} point out, the simple fixed fraction of Eddington model
fits the reverberation data very well but this model most clearly
contradicts the clustering results. If we are to reconcile the
clustering data with the reverberation data, we must conclude that there
must be some decoupling of  $M_{BH}$ and $M_{DMH}$  and this immediately
takes away the evidence for the flickering and any other single BH mass
model. Meanwhile, we postpone our discussion of reverberation  mapping
and the PLE model to Section 8 and Fig. \ref{fig:ple} where we shall
directly compare BH masses predicted by the PLE model to masses derived 
from reverberation mapping.

\section{Number densities of QSOs at all redshifts and luminosities.}

The space number density of QSOs compared to the numbers of remnant
nuclear black holes in early-type galaxies has been used to argue for
short ($\tau\approx10^{7}$yr) QSO lifetimes \cite{richstone}. However,
if the QSOs were restricted to a small range of host galaxy/halo/BH masses as
implied by the similarity of the QSO clustering results then this
argument would start to push QSO lifetimes higher. Generally, these
estimates depend on an assumption that QSO luminosity is proportional to
host BH mass and this assumption is also challenged by the lack of a
luminosity dependence of QSO clustering. \citet{richstone} suggest that
the bright QSO space density at $z\approx2$ is $\approx0.1$\% of the
galaxy density at $z=0$. Thus, for example,  if the QSOs are restricted
to $\approx1$\% of galaxies then this lifetime would increase to
$\approx10^9$yrs. Clearly the more restricted the galaxy luminosity
range, the longer these duty cycle arguments allow the QSO lifetime to
be.

More detailed arguments for short QSO lifetimes have been similarly made
on the basis of high  peaks models of biased QSO clustering.
\citet{martini} made predictions for the QSO correlation function based
on the halo number densities inferred  from  QSO lifetimes estimated
from QSO and model halo densities. Using the $\Lambda$CDM results from
these authors and the 2QZ clustering amplitude, \citet{croom05} derived
a QSO lifetime of 4-50 Myr at $z\approx2$. We note that a  monotonic
increase of halo mass with  QSO luminosity is an implicit assumption
here, otherwise it is impossible to establish a minimum halo mass at a
fixed QSO luminosity limit. In what follows we shall also make this
assumption. If we take the change in $b$ as $\approx9$\% from 2SLAQ
($r_0=5.7$h$^{-1}$Mpc) to SDSS QSOs ($r_0=6.3$h$^{-1}$Mpc) then from the
approximate relation, $b\propto M_{min}^{0.2}$, fitted to Fig. 6 of
\citet{martini}, this corresponds to an $\approx1.5\times$ bigger halo
mass for the $\Lambda$CDM model. Now assuming an approximate 
$\Lambda$CDM halo mass function with $N(<M)\propto M^{0.15}$ for
$M<10^{14}\rm{h}^{-1}M_\odot$ \citep{jenkins01} we find that the number
of haloes in the $2-3\times10^{12}\rm{h}^{-1}M_\odot$ mass range is
$\approx10$\% of the total with $M>2\times10^{12}\rm{h}^{-1}M_\odot$.
This means that the estimates of QSO lifetime would now lie in the range
$\approx40-500$ Myr.  Since the clustering of the 2SLAQ, 2QZ and SDSS
samples are statistically consistent, the increase in QSO lifetime could
even be larger than this estimate. 

This conclusion from the clustering is  stronger
than from the direct QSO imaging results. Here there is an $\approx2$mag
range in host luminosity or a factor of $\approx6$, corresponding to
$\approx40$\% of the  haloes with $M>2\times10^{12}M_\odot$
giving an $\approx2-3\times$ increase in QSO lifetime. Again, since the
range for the host luminosities is clearly an upper limit due to no
account being taken of galaxy magnitude errors, there is no necessary
disagreement with the clustering results and indeed the QSO lifetimes
could be much higher if as expected the galaxy magnitude  errors
dominate the $\approx2$mag scatter. We therefore suggest that the short
QSO lifetimes previously derived on the  basis of the \citet{martini}
arguments are at best lower limits if the QSOs are restricted to a range
of halo masses and even on the basis of the present clustering data
could imply that this lower limit is $\tau>10^8$yr.


\section{Long-lived QSO clustering model}
If QSO lifetimes become  longer then there is a clear prediction for QSO
clustering via  the long-lived model of \citet{fry96} (see also
\citet{cs96}). Here QSOs (and their host galaxies) are assumed to form
at fixed bias at high redshift and then only evolve as separate entities
clustering under gravity. This gives the relation:-

$$b_Q(z)=1+[b_Q(z=0)-1]/D(z)$$

\noindent where $D(z)$ is the gravitational growth factor and
$D(z)=1/(1+z)$ for the EdS model. This then produces the evolution for
$\xi_{20}$ shown in Fig. \ref{fig:xi20_lam}. The relation produces a
significantly worse fit than the single halo mass model discussed
earlier, falling with redshift whereas the data appears to increase.  If
we now include  the lower redshift results of \citet{georg} and
\citet{wake04} then the higher clustering implied at lower redshifts
produces improved agreement with the long-lived model prediction. We
find that the best fit bias is $b_Q(z=0)=1.42\pm0.03$. However, the
model is still rejected at high significance with $\chi^2=80.2$ for 27
data points (26 degrees of freedom). Note that following \citet{croom05}
this  bias for the long-lived model is uncorrected for non-linear effects.

Given that the model assumes a constant bias for all QSOs at formation,
this model intrinsically predicts a correlation function compatible with
a constant halo mass for QSOs independent of luminosity, at least at
fixed redshift.  However, it is true that if there was a strong
luminosity dependence at fixed $z$ shown by the QSO data then this might
also have been compatible with the long-lived model. 

We note that the cosmology also plays a role here. If we assume an EdS
cosmology then  $\xi_{20}$ reduces by a factor $\approx2$ by $z\approx2$. 
While the situation in respect of the single
halo mass model remains the same as for the $\Lambda$CDM case, the
situation in respect of the long-lived model improves significantly,
with the model now fitting the  high SDSS AGN clustering amplitude
while being able to fit better the  $\xi_{20}$ points at high redshift. The
best fit bias is $b_Q(z=0)=1.71\pm0.03$, assuming a $\tau$CDM
normalisation of $\bar{\xi}_{20}=0.12$. Overall the model is acceptable
in a chi-square test at the $\approx8$\% significance level
($\chi^2=36.7$  for 27 data points).

\citet{croom05} argued against a long-lived model on the grounds that
accretion must take place and the halo masses would change contradicting
the evidence from the constant mass clustering with redshift. However,
this argument is now weakened by the fact that if applied to early-type
galaxies and LRGs a contradiction  also results in the context of the
$\Lambda$CDM cosmology which is avoided by appealing to halo occupation
distributions (HODs). Indeed, the evolution of the luminosity function
and clustering of early-type galaxies themselves appear to be reasonably
compatible with long-lived PLE models. The LRG luminosity function
appears consistent with  passive evolution out to $z\approx0.7$
\citep{wake06}. Indeed, the NIR H and K galaxy number counts, dominated
by early-type galaxies at $K<20$, show little evidence of strong
evolution out to $z>1$ \cite{nm06, nm01, nm96}. At large scales,
\citet{wake08} found that for separations $r>1$h$^{-1}$Mpc, the
evolution of LRG clustering was reasonably consistent with the
long-lived, biased evolution out to $z\approx0.6$. At $r<1$h$^{-1}$Mpc
there was some evidence that the evolutionary model of \citet{fry96} was
too strong for the data. However, a HOD with a low galaxy merger rate of
only 2.4\%/Gyr can fit the data. \citet{sawangwitwtheta} also found that
a simple virialised cluster model,  which clearly also conserves galaxy
numbers, can improve the fit of a long-lived model to the small-scale
clustering at $r<1$h$^{-1}$Mpc. Although these results generally only
apply to $z<1$, the fit of a HOD model with {\it only a small amount of
galaxy merging but with a halo merging rate which is compatible with
$\Lambda$CDM} suggests that the simple arguments of \citet{croom05}
against similar long-lived models for QSOs may not be as strong as
initially envisaged.

\section{PLE Model predictions}

\begin{figure}
\includegraphics[scale=0.5]{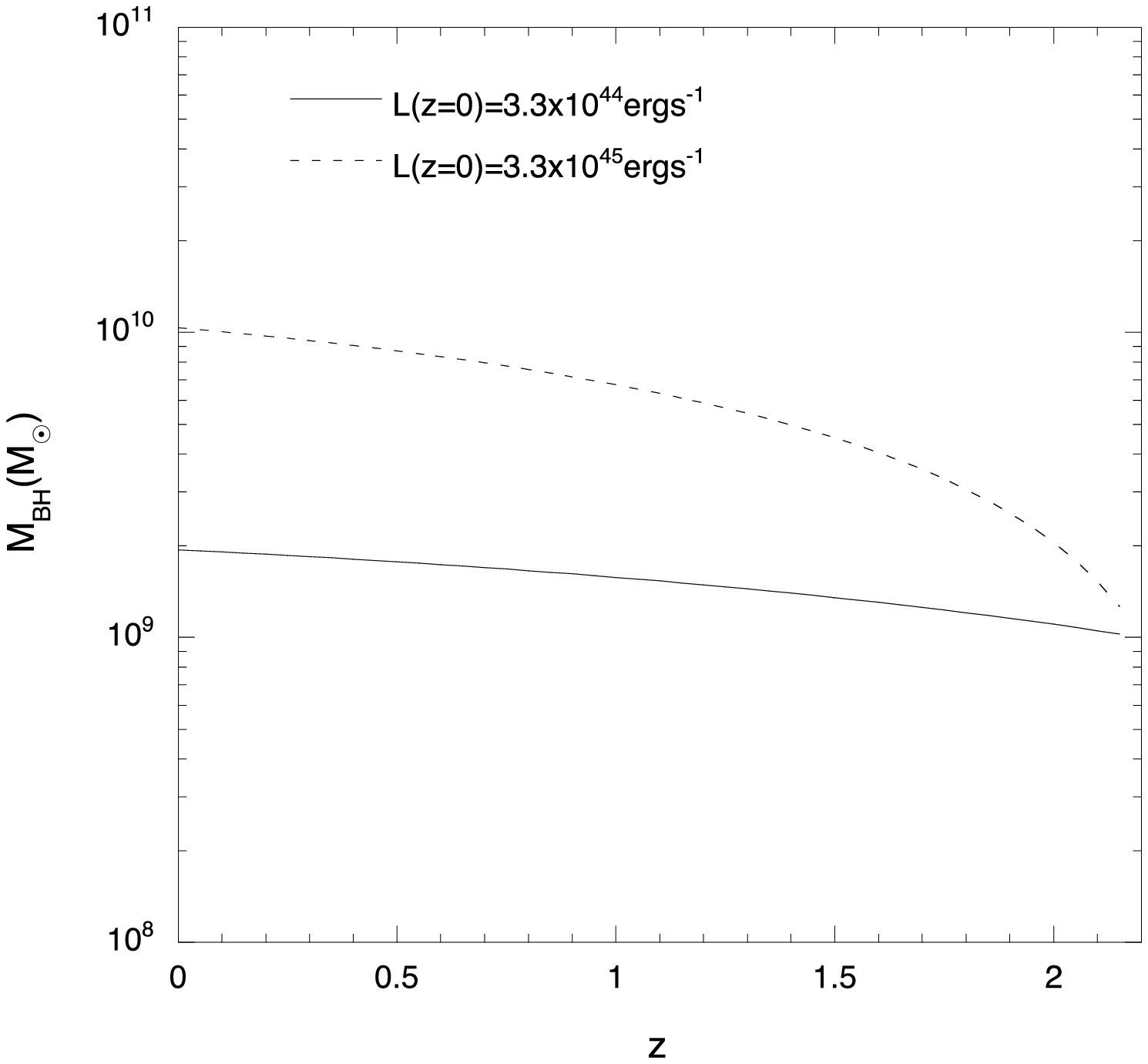}
\caption{$M_{BH}(z)$ for PLE  models for an $L^*$ QSO and  a $10L^*$
QSO, assuming $\gamma=3$, $\epsilon=0.4$ and $M_{BH}(z=2.2)=1\times10^9
M_\odot$. The $L^*$ model shows a relatively slow variation with $z$.
The $10L^*$ model shows a somewhat faster relation with $z$ and a
significant difference with the $L^*$ model at fixed $z$.}
 \label{fig:ple_gam_eps}
\end{figure}

We next connect the above clustering results to the major observation of
the evolution of the QSO luminosity function. One model which predicts a
`long-lived' evolution with redshift is a pure luminosity evolution
(PLE) model. If the dependence of QSO luminosity on redshift is given by
$L(z) = L(z=0) (1+z)^\gamma$, then the dependence of BH mass on redshift
is given by $$M_{BH}={3\over{2(\gamma-3/2)}}{L(z=0)\over \epsilon\times
c^2} t_0 (3.2^{\gamma-3/2}-(1+z)^{\gamma-3/2})$$ \noindent where
$\epsilon$ is the BH efficiency. At the characteristic $L^*$ knee in the
luminosity function, an $M^*_B(z=0)=-21.5-5{\rm log~h}_{70}$ mag QSO has
luminosity $L^*(z=0)=3.3\times10^{44}h_{70}^{-2}$ ergs$^{-1}$  in the
range 1200-10000\AA \citep{marshall84}. We have assumed the EdS model
for analytic simplicity but with fixed Universal age, $t_0=13.7$ Gyr.
For zero initial mass at $z=2.2$, $\epsilon=0.4$ and $\gamma=3$ the QSO
then accretes $M_{BH}\approx 10^9 M_\odot$ by $z=0$. We have seen that
the low redshift QSO clustering results may prefer such a long-lived
model. On the assumption that the halo and long-lived models are
self-consistent, the rise in predicted clustering amplitudes from
$z\approx2$ to $z\approx0$  implies an increase of $\approx10\times$ in
halo mass, since $b\propto M_{halo}^{0.2}$ from Section 6. Thus if the
halo and galaxy growth rates are not decoupled as assumed in the
previous section, then $M_{BH}$ would increase by a factor of
$\approx100\times$ (since $M_{BH}\propto M_{halo}^{1.85}$ from Section
3) and the PLE  model would be rejected on the basis of its much slower
predicted $M_{BH}$ growth rate (see Fig. \ref{fig:ple_gam_eps}). Even if
 the long-lived clustering model now gives an improved {\it statistical}
fit  with the inclusion of the low-z data in Figs. \ref{fig:xi20_lam},
the {\it theoretical} PLE $M_{BH}$ growth rate and the {\it theoretical} long-lived
model growth rate are mutually incompatible if $M_{BH}$ and $M_{DMH}$
follow the above relation.

\begin{figure}
\includegraphics[scale=0.5]{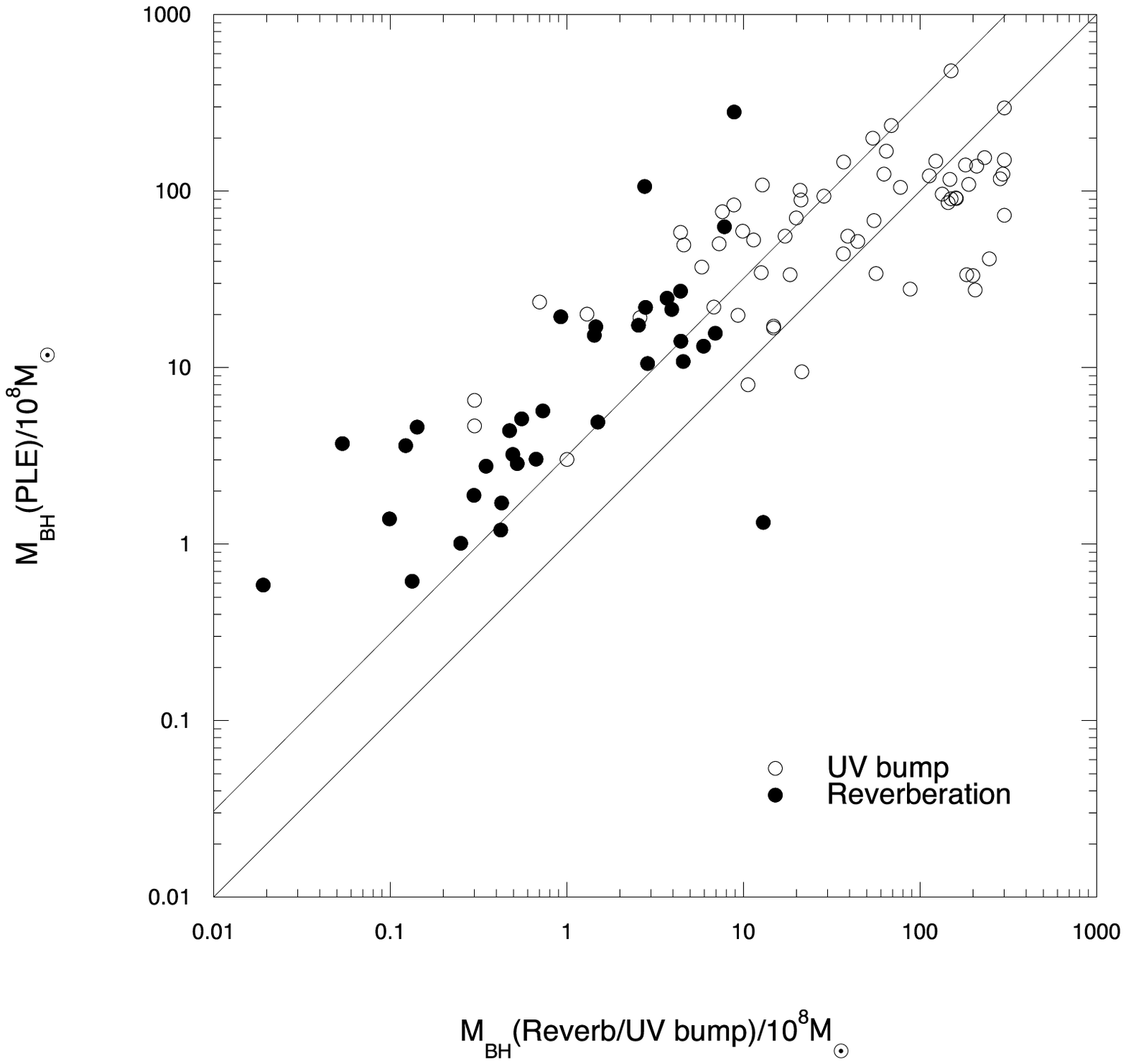}
\caption{PLE masses from the $M_{BH}(z)$ relation in the text compared
to the `UV-bump' fitted masses of \citet{laor} for 62 Palomar-Green
survey QSOs and reverberation masses of 35 QSOs from \citet{bmp04}. The
masses of \citet{laor} assume $h_{75}=1$ and $q_0=0.5$, similar to what
is assumed for the masses of \citet{bmp04} and for the PLE model. The
PLE model also assumes $\gamma=3$, $\epsilon=0.4$ and zero initial BH mass.
A line of equal mass and a line where the PLE masses are $3\times$ larger
than the other masses are shown.}
 \label{fig:ple}
\end{figure}

Fig. \ref{fig:ple} shows the  BH mass for 62 bright Palomar-Green QSOs
as estimated from fitting the `UV bump' in their spectra \citep{laor},
compared to the BH masses estimated from the PLE equation above. At the
high mass end, the PLE masses seem in good agreement with the masses of
\citet{laor}, although at low BH masses, the PLE masses appear too high.
Overall the agreement seems reasonable but this is the result for a PLE
model with negligible initial BH mass. A single BH mass model 
would depend  on the scatter around $\approx10^{9}M_\odot$
and the errors on the UV bump BH masses to explain the range of
BH mass shown by the \citet{laor} estimates in Fig. \ref{fig:ple}.

In a PLE model, the QSO is long-lived, thus explaining why the QSO space
density at $L^*$ remains the same with redshift \citep{marshall,
boyle88, croom04mn,croom09}. In any other model this represents a
coincidence. Basically the same luminosity function behaviour is also
seen in X-ray surveys \citep{boyle93}. There have been some  reports
that the X-ray LF is inconsistent with  PLE \citep{hasinger, giacconi,
alexander, worsley} but these deviations are generally small and at low
$L_X$ and in a  broad-brush way at least,  PLE remains an approximate
fit to the X-ray LF. PLE has always been an excellent phenomenological
fit to the optical QSO LF (although see \citealt{bongiorno}) but the question has been whether it has any physical
meaning. If exactly true the PLE model would suggest that low $z$
Seyfert I's would have as high a BH mass as high-z QSOs at fixed $L/L^*$
because the BH fuelling rate at low redshift is small compared to that
at high redshift \citep{marshall, boyle88}.

A single BH mass independent of QSO luminosity is not, however, a
prediction of PLE at fixed redshift. The equation above implies that for
fixed $\gamma$ and $\epsilon$ a QSO $10\times$ brighter than $L^*$ will
produce a $10\times$ bigger black hole mass of $10^{10}{\rm M}_\odot$ by
$z=0$. Indeed, for a $10L^*$ QSO, from $z=2.2$ to $z=1.5$,
of order the minimum QSO lifetime, a black-hole mass of $5\times10^9{\rm
M}_\odot$ will already have been created. If we assume that the QSO has
to be radiating at no more than Eddington, this implies that the initial
mass for such QSOs has to be $10^{9}{\rm M}_\odot$. If the fainter QSOs
have a similar initial mass then this will minimise any $M_{BH}$
difference between bright and faint QSOs (see Fig. \ref{fig:ple_gam_eps}). 

Of course, if halo and BH masses are completely decoupled, then there
may be no need to accommodate the approximate luminosity independence of
clustering in this way. However, it could be that although halo/BH mass
{\it growth rates} are decoupled, {\it initial}  halo and BH masses
could still be correlated. If so, the $M_{BH}:M_{halo}$ relation must
evolve with redshift in the sense that the halo mass containing a
particular $M_{BH}$ must decrease as $z$ increases. At $z=2.2$, all QSOs
would then start with the same initial $M_{BH}$ and have the same halo
mass to be consistent with the clustering observations. Higher
luminosity QSOs then accrete at a faster rate as in
Fig. \ref{fig:ple_gam_eps} but no luminosity dependence  of the QSO
clustering is generated because of the decoupling of halo and QSO/galaxy
merging rates and the clustering of all QSOs grows at the long-lived
rate.

In the context of the original PLE model,  QSOs at all luminosities at
$z\approx2$ could radiate at close to the Eddington limit and those at
lower redshift then would radiate at increasingly sub-Eddington rates,
up to $\approx30\times$. In the above PLE model case where all the
initial $M_{BH}$ must now be the same to fit the clustering results, if
we assume that at $z\approx2$ the brightest QSO is radiating at
Eddington, then at the same redshift the fainter QSOs have now  to
radiate at $\approx10\times$ lower than the Eddington rate. At
$z\approx0$, these intrinsically faint QSOs have thus now to be
radiating at $\approx300\times$ below Eddington. This may be  physically
feasible in that standard optically thick, physically thin, accretion
discs can accrete at down to $\approx1$\% of the Eddington rate (C.
Done, priv. comm.). Below this rate the accretion disc is unstable to
evaporation. Thus the -27 mag QSOs at high redshift and  the -22 mag
Seyfert I's locally can both be powered by thin disc accretion. Thus
there appears to be a reasonable range of QSO luminosities where the PLE
or single-BH mass models can  operate within the context of  a thin disc
accretion model.

The mean BH mass measured in low luminosity SDSS AGN by \citet{heckman}
via the $M_{BH}:\sigma$ relation is $\approx10^8 M_\odot$ and this seems
in less good agreement with the $\approx10^9 M_\odot$ PLE prediction
than the results of \cite{laor}. However, as we have already noted, the
SDSS AGN comprise some 18\% of the galaxy population whereas broad lined
QSOs comprise $<1$\%. Thus it is plausible that, although some fraction,
($<30$\%?) of the SDSS AGN may be obscured counterparts of QSOs, the
remainder might be expected to be a genuinely lower luminosity AGN
population, sufficiently sub-Eddington to make any thin accretion disk
unstable. Thus the obscured QSO population in a PLE model might lie only
in the high BH mass tail of the SDSS AGN mass distribution which extends
beyond $10^9 M_\odot$ (see Fig. 1 (left panel) of \citet{heckman}).

We also revisit the reverberation BH masses of \citet{bmp04} for 35 low
redshift QSOs and compare these with their  predicted PLE masses as we
did for the `UV-bump' masses of \citet{laor} (see Fig. \ref{fig:ple}).
We see that there is a good  correlation but that the reverberation
masses are generally $\approx5\times$ higher than the PLE predictions
and the agreement is  less good than for the masses of \citet{laor}. For
the 11 QSOs in common between the two samples, the `UV-bump' masses are
$\approx3\times$ higher than the reverberation masses but there is a
significant scatter. The amplitude of the reverberation masses contains
a scaling factor which was adjusted to fit the $M_{BH}:\sigma^*$
relation in quiescent galaxies \citep{bmp04}. We also note that if
we chose to assume a $\approx10^9 M_\odot$ initial BH mass then the
disagreement would clearly worsen between the PLE and reverberation
masses. Nevertheless, the correlation with the reverberation masses seen
in Fig. \ref{fig:ple}  means that the PLE model performs better in
this respect than  if the single BH mass  or flickering models were  assumed to
apply at $z\approx0$  as at $z>0.5$ (see Section 5).

\section{QSO X-ray variability}

A traditional  problem for the PLE model
is whether the massive BH it predicts, particularly in low redshift
Seyfert I's, can accommodate the short timescale variability observed in
some of these sources.  The  timescale of variability set by a
$\approx10^{9}M_\odot$ BH mass is $\approx10^4$s or $\approx3$hrs. Now
variability in many QSOs is seen on timescales of $<1$hr and the
question is whether these can be localised flares, affecting less than
5-10\% of the area and total luminosity of the QSO. Clearly the smaller
the fraction of the QSO luminosity that the flares affect, the more
plausible the argument for the high PLE BH mass becomes. But   
the fact that the optical and X-ray evolutions are  similar
is a further coincidence that a long-lived model like PLE explains,
since the optical accretion disc is expected to heat the X-ray `corona'.
We now consider these issues in more detail.

The PLE and single BH mass  models both  make simple predictions for QSO
variability. PLE predicts that at fixed $L/L^*$, the amplitude of
variability will be approximately independent of redshift because the BH
mass only changes slowly with redshift. The single BH mass model makes
the even simpler prediction that  the variability amplitude will be
approximately independent of redshift {\it and} luminosity, i.e. $L/L^*$
or $L$. \citet{almaini} found in a statistical X-ray variability
analysis of 86 QSOs from  the Deep ROSAT QSO survey \cite{georg96} that
generally the variability of QSOs shows a relatively flat distribution
with redshift (see their Fig. 6b). They also found that there was some
evidence for an inverse correlation with $L_X$ at $z<0.5$ which was
similar to that found for low-z AGN by \citet{lawrence, nandra} but this
correlation disappeared for $z>0.5$ (see their Figs. 8a,b). More
recently, other authors have made similar studies of QSOs in XMM and
Chandra observations. In the CDF-S, \citet{paolillo} analysed 74
variable QSOs and found little evidence of dependence on $L_X$ (see
their Fig. 12). In the XMM-Newton Deep Survey of the Lockman Hole,
\citet{mateos} found variances for 74 QSOs and AGN that were
approximately independent of luminosity and redshift (see their Fig. 5).
Finally, \citet{papadakis} in the same XMM-Newton Lockman Hole QSO
dataset found no significant correlations with either redshift or
luminosity (see their Fig. 2). We conclude that while at $z<0.5$, there
seems evidence for an inverse correlation between luminosity and
variability amplitude, at $z>0.5$ there seems little evidence of any
variability correlation with luminosity or redshift. While the latter
results argue for both the PLE and single BH mass model, the former
result would only be consistent with the PLE hypothesis. Finally, we
note that the amplitude of QSO variability is in the range
$\sigma\approx0.1-0.3$. Thus this is in the range 10-30\% of flux at
which level it may be easier to argue that the high BH masses predicted
by PLE models may be allowed, if only a fraction of the QSO luminosity
is involved in the variability.

\section{Discussion}

SDSS, 2QZ and 2SLAQ QSOs show  similar clustering scale lengths to
within $\approx\pm10$\%, while spanning an order of magnitude in
luminosity. The QSOs in these surveys also show slow clustering
evolution with redshift and a model with a single halo/BH  mass for all
QSOS fits both these results well. If true, then it implies that if the
most luminous SDSS QSOs are radiating at Eddington luminosities at
$z\approx2$, then the  faintest 2SLAQ QSOs at lower redshift  must be
radiating at an $\approx100\times$ lower rate than Eddington. Although
problems with BH masses estimated from emission-line widths can be
avoided by appealing to  evolution in the BLR radius-luminosity relation
or in the line-widths themselves, it may be more difficult to explain
away the strong $M_{BH}-L$ correlation implied at $z=0$ from
reverberation mapping. This presents an immediate problem for any single
BH mass model and variants designed to fit QSO clustering such as the
`flickering' model  of \citet{lidz06}.

If QSO halo mass is restricted to a very small range then estimates of
QSO lifetime must  rise significantly and our very rough estimate is by
a factor of $\approx40$ to $\approx10^9$ yrs based on the arguments of
\citet{martini}. This then  prompts consideration of the clustering
prediction of a long-lived QSO model \citep{fry96}. We find that this
model predicts an increase in clustering amplitude at $z<0.5$. When we
include the low-redshift AGN clustering results of \citet{wake04} and
\citet{georg}, we find improved  agreement with the long-lived model.
Including these results  also tends to worsen the fit of the single mass
halo model. If the QSOs are long-lived then this also tends to argue against the
idea that QSOs always radiate at a fixed fraction of $L_{Edd}$ because 
QSO BH mass tends to increase with time while QSO luminosity drops, so even if 
QSOs start at the Eddington rate they quickly become sub-Eddington.

If the  long-lived QSO model is correct, then the QSO clustering results
may be consistent with the fit of PLE models to the QSO luminosity
function. A physical interpretation of why the QSO space  density at
$L^*$ is observed to remain approximately  constant from $z=2$ to $z=0$
is that the QSOs have a long lifetime and  the luminous Type I QSOs seen
at $z=2.2$ dim by a factor of $\approx30$ to become the low-luminosity
Seyfert I's seen at z=0. Thus this PLE model prediction may be supported
by the QSO clustering results. We have noted that  if $M_{BH}$ and
$M_{DMH}$ follow relations such as $M_{BH}\propto M_{DMH}^{1.8}$, then
the {\it theoretical} growth rates of $M_{BH}$ from PLE and the
long-lived model are mutually incompatible. However, a key feature
of the long-lived model is that it implicitly assumes that galaxy/BH 
and halo mass growth rates are decoupled so that the  BH mass growth
rate with $z$ predicted by PLE is easily accommodated. At fixed $z$, PLE
also predicts that  BH mass growth rates  are proportional to QSO
luminosity. But again, since in the context of long-lived models, halo
and BH mass growth are naturally  decoupled there is no necessary
contradiction with the clustering results here for PLE. This  decoupling
has given us the freedom to assume a constant initial BH mass for the
PLE model so that it is actually in line with the luminosity independent
clustering results at high $z$. Then the higher accretion rate of the
more luminous QSOs gives the possibility of generating the luminosity-BH
reverberation mass correlation found at $z=0$, without also generating
clustering that is too luminosity dependent at these redshifts. In this
picture, it is the lack of a relation between halo and galaxy/BH mass
that explains the lack of luminosity dependence of QSO clustering seen
in Fig. \ref{fig:xis}, rather than the lack of a relation between
luminosity and BH mass.

Recently, \citet{korm11a,korm11b} have presented evidence that
nuclear velocity dispersions may not correlate well with galaxy circular
velocities in cases they have observed where there is no bulge, just
nuclear star-clusters. They also argue that dynamically measured
$M_{BH}$  do not correlate well with disc stellar masses. Since circular
velocities and disk stellar masses usually  correlate well with DM halo
masses, they then conclude that DMH masses don't correlate well with
$M_{BH}$. These results  have some resonance with the arguments we have
presented above based on QSO clustering and also the direct imaging of QSO
hosts. We have argued that these results imply either that
QSOs have the same BH mass or that the halo and BH mass growth rates are
decoupled. In particular, our long-lived PLE model argues for a
decoupling of halo and BH mass at least at $z<2.2$ to reconcile the slow
evolution of QSO clustering with the faster predicted evolution of halo
clustering. In the models of \citet{lidz06} it is the QSO luminosity
that decouples from the BH/DMH masses to explain the clustering results.
But in these scenarios the BH and DMH masses are expected to be strongly
correlated, in contradiction with the new results of \citet{korm11a}.

\section {Conclusions}
For clarity, we  summarise  the arguments for and against the four  QSO 
models we have discussed above:

\renewcommand{\labelitemii}{$-$}

\begin{itemize}
\item QSOs radiate at a fixed fraction of  $L_{Edd}$:

-in favour - reverberation mapping $M_{BH}-L$ relation

           - agrees with virial estimates of $M_{BH}$ using unevolved BLR $L-r$ relation. 

- against - this model predicts strongly luminosity dependent clustering which 
is not observed; need to break the $M_{BH} - M_{DMH}$ relation as suggested by
\citet{korm11a}.
        
          - if the QSOs are long-lived as implied by the luminosity independence of the
          clustering then they cannot always operate at a fixed fraction of $L_{Edd}$
          and be consistent with observed QSO luminosity evolution.

\medskip

\item The `flickering' model of \citet{lidz06}:

- in favour - $L$ independent $z>0.5$ QSO clustering.

- against - the reverberation mapping $M_{BH}-L$ relation at $z\approx0$.
          
          - virial estimates of $M_{BH}$ using the unevolved BLR $L-r$ relation. 

          - the longer duty cycles suggested by the coherent QSO LF
            evolution.

          -  \citet{korm11a} evidence against the assumed  $M_{BH} - M_{DMH}$ relation.

\medskip

\item   The single BH mass model:

- in favour - $L$ independent $z>0.5$ QSO clustering. 

- against   - the rise in clustering amplitude at $z<0.5$.

            - the reverberation mapping $M_{BH}-L$ relation at $z\approx0$.
            
            - virial estimates of $M_{BH}$ using the unevolved BLR radius-luminosity 
            relation.
\medskip

\item The long-lived/PLE model:

- in favour - $L$ independence of QSO clustering implies long lifetime.

            - not ruled out by the QSO clustering evolution over the $0<z<2.2$ range

            - reasonably reproduces reverberation and UV bump $M_{BH}$.

            - explains coherent QSO LF evolution.

            - invokes a decoupled $M_{BH} - M_{DMH}$ relation as evidenced 
            by \citet{korm11a}.
    
- against    - virial estimates of $M_{BH}$ using the unevolved BLR $L-r$ relation.

             - short X-ray variability timescales for low-z Seyfert I.

\end{itemize}

It is clear that models which predict well  the QSO clustering results
then tend to have a problem with the reverberation and UV bump
$M_{BH}-L$  relation and vice versa. We take the view that it is easier
to circumvent the clustering results (by decoupling $M_{BH}$ and
$M_{halo}$) than the reverberation mapping results. Therefore the answer
to the question posed in the title may be that QSOs do not all have the
same black-hole mass, although they may occupy similar  halo masses at
fixed redshift. Overall, we conclude that a model that is reasonably
consistent with both clustering and the reverberation/UV-bump masses is
the long-lived PLE model.

\section*{Acknowledgments}

We thank an anonymous referee for comments which improved the 
quality of this paper.

US acknowledges financial support from the Institute for the
Promotion of Teaching Science and Technology (IPST) of
The Royal Thai Government. 

We  thank all the present and former staff of the 
Anglo-Australian Observatory for their work in building and operating
the 2dF facility. 

Funding for the SDSS and SDSS-II has been provided by the Alfred
P. Sloan Foundation, the Participating Institutions, the National
Science Foundation, the U.S. Department of Energy, the National
Aeronautics and Space Administration, the Japanese Monbukagakusho, the
Max Planck Society, and the Higher Education Funding Council for
England. The SDSS Web Site is {\tt http://www.sdss.org/}.

\setlength{\bibhang}{2.0em}
\setlength\labelwidth{0.0em}

\bibliographystyle{mn2e}
\bibliography{qso_bh.bib}

\label{lastpage}

\end{document}